\documentclass[aps,pra,eqsecnum,twocolumn]{revtex4}
\usepackage{graphicx}
\usepackage{bm}
\usepackage{amsmath,amssymb,amsthm,dsfont,bm,enumitem}
\usepackage{color}
\usepackage{wasysym}
\usepackage{ulem}
\usepackage[dvipsnames]{xcolor}

\usepackage[applemac]{inputenc}

\usepackage[colorlinks=true,urlcolor=blue,citecolor=blue,linkcolor=blue]{hyperref}

\begin{document}
\preprint{}
\title{Spatial and temporal coherence via polarization mutual coherence function} 
\author{Alfredo Luis}
\email{alluis@fis.ucm.es}
\homepage{https://sites.google.com/ucm.es/alfredo/inicio}
\affiliation{Departamento de \'{O}ptica, Facultad de Ciencias
F\'{\i}sicas, Universidad Complutense, 28040 Madrid, Spain}
\date{\today}

\begin{abstract}
We address polarization coherence in terms of correlations of Stokes variables. We  develop an scalar polarization mutual coherence function that allows us to define a polarization coherence time. We find a suitable spectral polarization density allowing a polarization version of the Wiener-Khintchine theorem. With these tools we also address the polarization version of the van Cittert-Zernike theorem. 
\end{abstract}
\maketitle

\section{Introduction}
Coherence is a fundamental physical concept at the hearth of classical optics and quantum physics \cite{MW95,EW07}. Moreover, coherence has been acknowledged in quantum theory as the actual resource for the emerging quantum technologies \cite{SP17,CG19}. 

Being such a fundamental principle it only manifest indirectly through some other observable phenomena. The standard realm where coherence is addressed is interference. However, another equally valid domain is polarization, which is actually simpler, more robust, and far more easier to handle than interference \cite{EW07,GO22,CB98}.

Polarization is conveniently expressed by the Stokes parameters, which involve correlations of complex-field amplitudes. In this work we go beyond to investigate polarization coherence in terms of correlations of Stokes variables. 

Following the works in Refs. \cite{SSKF08,SRFS17,SSKF09} we develop an scalar polarization mutual coherence function that allows us to derive a coherence time and an spectral polarization density. With these tools we address the polarization versions of two celebrated theorems in classical-optics coherence. These are the Wiener-Khintchine \cite{NW30,AK34,AY87} and van Cittert-Zernike \cite{vC34,FZ38} theorems , dealing with the time-frequency and spatial manifestations of coherence, respectively. Autocorrelation functions of the Stokes parameters and their propagation have been already addressed in the literature as a mean for studying the polarization-related spatial structure of light beams \cite{KV19,SPBS21,WHT23}.

\bigskip

\section{Polarization mutual coherence function}

Let us consider purely transverse fields fully described by a two-component complex vector 
\begin{equation}
\bm{E} (\bm{r},t) = \begin{pmatrix} E_x (\bm{r},t) \\  E_y (\bm{r},t) \end{pmatrix} ,
\end{equation}
whose polarization state is usually expressed in terms of the four Stokes parameters $\langle S_j  (\bm{r},t) \rangle$ as the ensemble average $\langle \bullet \rangle$ of the Stokes variables 
\begin{equation}
\label{Sv}
    S_j (\bm{r},t) = \bm{E}^\dagger (\bm{r},t) \boldsymbol{\sigma}^{(j)} \bm{E} (\bm{r},t), 
\end{equation}
where $\boldsymbol{\sigma}^{(j)}$, $j=0,1,2,3$ are the Pauli matrices, being $\boldsymbol{\sigma}^{(0)}$ the identity, and the superscript $\dagger$ represents Hermitian conjugation. The Stokes-variables vector will be defined by the last three components
\begin{equation}
\label{3D}
    \bm{S} (\bm{r},t) = \left ( S_1 (\bm{r},t), S_2 (\bm{r},t), S_3 (\bm{r},t) \right )^T, 
\end{equation}
where the superscript $T$ denotes transposition. Nevertheless, in the last section of this work we will re-derive the results considering the complete four-dimensional Stokes vector. 

\bigskip

With this we can introduce the polarization mutual coherence function, following the works in Refs. \cite{SSKF08,SRFS17,SSKF09}
\begin{equation}
\label{pcf}
    \Gamma_S (\bm{r}_1, \bm{r}_2, t_1, t_2 ) = \left \langle \bm{S} (\bm{r}_1, t_1) \cdot  \bm{S} (\bm{r}_2, t_2) \right \rangle ,
\end{equation}
as a convenient generalization of the standard scalar-field coherence function,
\begin{equation}
    \Gamma_E (\bm{r}_1, \bm{r}_2, t_1, t_2 ) = \left \langle E^\ast(\bm{r}_1, t_1) E (\bm{r}_1, t_2) \right \rangle .
\end{equation}

\section{Polarization Wiener--Khintchine theorem}

In this section we will focus on the temporal dependence of $\Gamma_S$, so for the sake of simplicity  we shall omit the spatial dependence. Let us assume that the process is stationary so that $ \Gamma_S ( t_1, t_2 ) $ is invariant under time translations of the same amount on both time arguments $ \Gamma_S ( t_1 +t, t_2 + t ) = \Gamma_S ( t_1, t_2 ) $. This implies that $\Gamma_S$  must depend on time just through the time difference $\tau= t_2-t_1$, this is $\Gamma_S ( t_1, t_2 ) = \Gamma_S (\tau)$.

\bigskip

Let us consider an spectral decomposition of the Stokes variables $\bm{S} (t)$ in the form 
\begin{equation}
    \bm{S} (t) = \int d \Omega \bm{S} (\Omega) e^{- i \Omega t}, \quad \bm{S} (\Omega) = \bm{S}^\ast (- \Omega) ,
\end{equation}
where the asterisk denotes complex conjugation, so the last conditions holds to ensure their real character. With this we get 
\begin{equation}
    \Gamma_S (t_1 ,t_2 ) = \int d \Omega d\Omega^\prime 
\left \langle  \bm{S}^\ast (\Omega) \cdot   \bm{S} (\Omega^\prime) \right \rangle e^{i \Omega t_1} e^{ -i \Omega^\prime t_2} .
\end{equation}
Stationarity demands that 
\begin{equation}
    \left \langle  \bm{S}^\ast (\Omega) \cdot   \bm{S} (\Omega^\prime) \right \rangle \propto \delta (\Omega - \Omega^\prime) ,
\end{equation}
so that 
\begin{equation}
    \Gamma_S (\tau ) = \int d \Omega
\left \langle  | \bm{S} (\Omega) |^2 \right \rangle e^{- i \Omega \tau} .
\end{equation}
This has the form of a polarization Wiener--Khintchine theorem expressing that the polarization mutual coherence function $\Gamma_S (\tau )$ is the Fourier transform of a so defined spectral polarization density $\left \langle  | \bm{S} (\Omega) |^2 \right \rangle $. 

\bigskip

We may pose the corresponding Fourier inversion of this relation
\begin{equation}
\left \langle | \bm{S} (\Omega) |^2 \right \rangle = \frac{1}{2 \pi} \int d \tau   \Gamma_S (\tau ) e^{ i \Omega \tau} .
\end{equation}
However this requires some bit of caution. As it has been well noticed in Ref. \cite{SSKF08} when there is partial or total polarization it holds that $\Gamma_S (\tau)$ does not tend to zero as $\tau$ tens to infinity. {\it Grosso modo}, this is because for stationary partially polarized fields the Stokes parameters are not zero and independent of time, maintaining polarization correlation forever. To some extend this spoils the inversion of the Fourier transform. More specifically the polarized part of the  beam would be reflected in a Dirac delta type contribution to $\left \langle | \bm{S} (\Omega) |^2 \right \rangle $ at $\Omega=0$, that impedes a suitable assessment of coherence time for example. 

\bigskip

These difficulties can be avoided by removing from the analysis the fully polarized part of the field, that actually requires no statistical description. This can be done focusing directly on polarization fluctuations defining a mutual coherence function for polarization fluctuations, always in the stationary case, 
\begin{equation}
    \Gamma_{\delta S} ( \tau ) = \left \langle \delta \bm{S} (t) \cdot  \delta \bm{S} (t + \tau) \right \rangle ,
\end{equation}
where 
\begin{equation}
\label{pf}
    \delta \bm{S} ( t) = \bm{S} ( t)  - \left \langle \bm{S} (t)  \right \rangle .
\end{equation}
We may say that this is a particular form of Hanbury Brown-Twiss correlations, written in a slightly different way than usual. This is because deep down Stokes parameters at a single point are intensities \cite{MW95}. Actually, intensity correlations may be included explicitly in the analysis through the first Stokes variable $S_0$, as we will examine in the last part of the work where we consider a mutual coherence function involving the four Stokes parameters.

\bigskip

This allows us to provide the following  Wiener--Khintchine theorem for polarization fluctuations expressed by the following pair of Fourier transforms
\begin{equation}
\label{pWKt}
    \Gamma_{\delta S}  (\tau ) = \int d \Omega
\left \langle  | \delta \bm{S} (\Omega) |^2 \right \rangle e^{- i \Omega \tau} ,
\end{equation}
and
\begin{equation}
\label{pWKt2}
\left \langle | \delta \bm{S} (\Omega) |^2 \right \rangle = \frac{1}{2 \pi} \int d \tau  \Gamma_{\delta S}  (\tau ) e^{ i \Omega \tau} .
\end{equation}

\bigskip

This approach then provides us with a suitable definition of a finite polarization-fluctuations coherence time in a form already used in scalar-field coherence as \cite{Perina}
\begin{equation}
\label{tauc}
    \tau_c = \frac{1}{\Gamma_{\delta S} ^2 (0)}\int d \tau \Gamma_{\delta S}^2 (\tau ) .
\end{equation}
This differs from previous definitions of polarization time such as the one in Ref. \cite{SSKF08}, as discussed in more detail when considering some particular examples.

As a further property of this approach we have that this  polarization time $\tau_c$ enters in a suitable exact duality relation with the the  width $\Delta \Omega$ of $ \langle | \delta \bm{S} (\Omega) |^2 \rangle $ defined in terms of R\'{e}nyi entropy \cite{Renyi} as
\begin{equation}
    \Delta \Omega = \frac{1}{2\pi} \frac{\left [ \int d \Omega^\prime \left \langle  | \delta \bm{S} (\Omega^\prime) |^2 \right \rangle \right ]^2 }{\int d \Omega \left \langle  | \delta \bm{S} (\Omega) |^2 \right \rangle^2 } ,
\end{equation}
which follows form the Parseval's theorem as
\begin{equation}
    \tau_c \Delta \Omega = 1 .
\end{equation}

\bigskip

We have considered no assumption about the field statistic other than being stationary. Otherwise the field statistics is quite general and the relations found hold in any case. However, in order to proceed addressing meaningful practical situations we can consider in more detail the usual case of Gaussian statistics. Nevertheless, we will explore somewhat beyond that in the last section of the paper.

\bigskip

\subsection{Gaussian statistics}

In order to address Gaussian statistics we can begin with the polarization mutual coherence function  
\begin{equation}
\Gamma_S (t, t + \tau ) = \langle \bm{S} (t) \cdot  \bm{S} (t + \tau ) \rangle  ,
\end{equation}
that  after Eq. (\ref{Sv}) 
\begin{equation}
\Gamma_S (t, t + \tau )  = \sum_j \langle \bm{E}^\dagger (t) \boldsymbol{\sigma}^{(j)} \bm{E} (t) \bm{E}^\dagger (t+\tau) \boldsymbol{\sigma}^{(j)} \bm{E} (t+ \tau ) \rangle, 
\end{equation}
for $j=1,2,3$, which can be expressed as
\begin{eqnarray}
& \Gamma_S (t, t + \tau ) = \sum_{j} \sum_{k, \ell,m,n} \sigma_{k,\ell}^{(j)} \sigma_{m,n}^{(j)} & \nonumber \\
 & & \nonumber \\
 & \times \left \langle E_k^\ast (t) E_\ell (t) E^\ast_m (t+\tau) E_n (t+ \tau ) \right \rangle , & 
\end{eqnarray}
for $k, \ell,m,n=x,y$. For Gaussian statistics, the fourth-order correlations can be expressed in terms of second-order ones via the Gaussian-moment theorem for complex variables, assuming zero means for simplicity, as \cite{MW95}
\begin{eqnarray}
& \Gamma_S (t, t + \tau ) = \sum_{j} \sum_{k, \ell,m,n} \sigma_{k,\ell}^{(j)} \sigma_{m,n}^{(j)} & \nonumber \\
 & & \nonumber \\
 & \times \left [ \left \langle E_k^\ast (t) E_\ell (t) \right \rangle \left \langle  E^\ast_m (t+\tau) E_n (t+ \tau ) \right \rangle \right . & \nonumber \\ & & \nonumber \\
  &+ \left . \left \langle E_k^\ast (t) E_n (t+ \tau )\right \rangle \left \langle  E^\ast_m (t+\tau)  E_\ell (t) \right \rangle \right ] . &
 \end{eqnarray}
In the first factor we recognize the product of Stokes parameters at two different times $ \langle \bm{S} (t) \rangle \cdot \langle \bm{S} (t + \tau ) \rangle$. This will be embodied in the mutual coherence function for polarization fluctuations, passing from $\Gamma_S$ to $\Gamma_{\delta S}$. For the second factor we may use the following property of Pauli matrices, that can be easily demonstrated by direct computation
\begin{equation}
\label{Pmp}
    \sum_{j=1,2,3} \sigma_{k,\ell}^{(j)} \sigma_{m,n}^{(j)} = 2 \delta_{k,n}\delta_{\ell,m}-\delta_{k,\ell}\delta_{m,n} ,
\end{equation}
that leads to 
\begin{eqnarray}
& \Gamma_{\delta S} (t, t + \tau ) = & \nonumber \\
 & & \nonumber \\
 & 2 \sum_{k, \ell}  \left \langle E_k^\ast (t) E_k (t+ \tau ) \right \rangle \left \langle  E^\ast_\ell (t+\tau)  E_\ell (t) \right \rangle & \nonumber \\
& & \nonumber \\
& - \sum_{k, m} \left \langle E_k^\ast (t) E_m (t+ \tau ) \right \rangle \left \langle  E^\ast_m (t+\tau)  E_k (t) \right \rangle , &
 \end{eqnarray}
this is to say
\begin{eqnarray}
\label{GdStW}
& \Gamma_{\delta S}  (t,t+\tau ) = 2  \left | \mathrm{tr} \bm{W} (t,t+\tau ) \right |^2 & \nonumber \\
& & \nonumber \\
 & - \mathrm{tr} \left [ \bm{W} (t,t+\tau )\bm{W}^\dagger (t,t+\tau ) \right ]  ,& 
\end{eqnarray}
where  $\bm{W} (t,t+\tau )$ is the $2 \times 2$ mutual coherence matrix
\begin{equation}
\bm{W} (t,t+\tau ) =  \left \langle \bm{E}^\ast (t)  \bm{E}^T (t + \tau )\right \rangle .
\end{equation}

\bigskip

\subsection{Two-time polarization}

We may write the result (\ref{GdStW}) in an slightly different form by expressing $\Gamma_{\delta S}$ as 
\begin{equation}
    \Gamma_{\delta S}  (t,t+\tau ) = \frac{1}{2} \left [ 3- \mathcal{P}^2 (t,t+\tau ) \right ] \left |\mathcal{S}_0 (t,t+\tau ) \right |^2 ,
\end{equation}
where the four-dimensional complex vector $\mathcal{\bm{S}}$ is defined as
\begin{equation}
\label{ttSp}
    \mathcal{S}_j (t,t+\tau ) = \mathrm{tr} \left [ \bm{W} (t,t+\tau ) \boldsymbol{\sigma}^{(j)} \right ] ,
\end{equation}
which may be termed two-time Stokes parameters, in analogy with the two-point Stokes parameters \cite{ED04,KW05}, and
\begin{equation}
    \mathcal{P}(t,t+\tau ) = \frac{|\bm{\mathcal{S}}( t,t+\tau )| }{|\mathcal{S}_0 (t,t+\tau )  |} ,
\end{equation}
is a time analog of the degree of cross polarization, where 
\begin{equation}
    \bm{\mathcal{S}}( t,t+\tau ) = \left ( \mathcal{S}_1( t,t+\tau ), \mathcal{S}_2( t,t+\tau ), \mathcal{S}_3 ( t,t+\tau ) \right )^T .
\end{equation}

\bigskip

\subsection{Gaussian stationary fields}

If the field is Gaussian and stationary then $\bm{W} (t,t+\tau )= \bm{W} (\tau )$ and we have 
\begin{equation}
\label{GdSt}
\Gamma_{\delta S}  (\tau ) = 2  \left | \mathrm{tr} \bm{W} (\tau ) \right |^2  - \mathrm{tr} \left [ \bm{W} (\tau )\bm{W}^\dagger (\tau ) \right ]  ,
\end{equation}
in full agreement with the results in Ref. \cite{SSKF08}. The two factors in Eq. (\ref{GdSt}) have clear meanings. Up to normalization factors, the first one is the Karczewski--Wolf degree of coherence for partially coherent fields \cite{BK63,EW03}, while the second one is the alternative definition introduced by Tervo, Set\"al\"a, and Friberg \cite{TSF03,STF04}. 

\bigskip

We think that the concurrence in Eq. (\ref{GdSt}) of two previously introduced alternative definitions of the degree of coherence is a fortunate fact that may help to better understand this complex and rather controversial issue of coherence in the polarization domain   \cite{BK63,EW03}.

\bigskip

To deepen in the stationary case via Fourier analysis let us consider the spectral decomposition of complex-field amplitudes
\begin{equation}
\label{sd}
    E_j (t) = \int d \omega  E_j (\omega ) e^{- i \omega t} ,
\end{equation}
where stationarity implies incoherent spectral components 
\begin{equation}
\label{si}
    \langle E_j^\ast (\omega ) E_k (\omega^\prime ) \rangle \propto \delta (\omega - \omega^\prime ) , 
\end{equation}
so that 
\begin{equation}
    \langle E_j^\ast (t ) E_k (t+\tau ) \rangle = \int d \omega \langle E_j^\ast (\omega ) E_k (\omega ) \rangle e^{-i \omega \tau} ,
\end{equation}
and then 
\begin{equation}
    \bm{W} (\tau )= \int d \omega \bm{W} (\omega )e^{-i\omega \tau} ,
\end{equation}
where $\bm{W} (\omega)$  the Hermitian cross-spectral density matrix
\begin{equation}
 \bm{W} (\omega) = \left \langle \bm{E}^\ast (\omega) \bm{E}^T (\omega) \right \rangle , \quad \bm{W}^\dagger (\omega) = \bm{W} (\omega) .
\end{equation}
Therefore after some simple algebra we get that the polarization Wiener--Khintchine theorem holds with
\begin{eqnarray}
\label{dSOm}
&  \langle \left |  \delta \bm{S} ( \Omega ) \right |^2 \rangle = \int d\omega \left \{ \right . 2 I(\omega)I(\omega- \Omega)& \nonumber \\
& & \nonumber \\
& -  \left . \mathrm{tr} \left [ \bm{W} (\omega) \bm{W} (\omega - \Omega) \right ]  \right \} , &
\end{eqnarray}
where $I(\omega) = \mathrm{tr}\bm{W} (\omega)$ is the usual spectral density. 

\bigskip

\section*{Example 1}

Let us illustrate this approach with two suitable examples taken from Ref. \cite{SSKF08}. The first one is a uniformly partially polarized stationary Gaussian beam with electric-field mutual coherence matrix 
\begin{equation}
    \bm{W} (t,t+\tau ) = \bm{J} e^{-\tau^2/(2 \sigma^2)},
\end{equation}
where the matrix $\bm{J}$ is constant. Then, taking advantage of its Gaussian character it can be readily seen that
\begin{equation}
   \Gamma_{ \delta S} (\tau) = \frac{I^2}{2} \left ( 3 - P^2 \right ) e^{-\tau^2/\sigma^2} ,
\end{equation}
where $I$ is the intensity and $P$ the degree of polarization 
\begin{equation}
    I = \mathrm{tr} \bm{J} , \quad P^2 = 2 \frac{\mathrm{tr} \bm{J}^2}{\mathrm{tr}^2 \bm{J}}- 1 .
\end{equation}
The spectral density of polarization fluctuations, either from Eq. (\ref{pWKt2}) or (\ref{dSOm}), is 
\begin{equation}
    \left \langle  | \delta \bm{S} (\Omega) |^2 \right \rangle = \frac{I^2}{4 \sqrt{\pi}}\left ( 3 - P^2 \right ) \sigma e^{- \sigma^2 \Omega^2/4} .
\end{equation}
The corresponding coherence time $\tau_c$ in Eq. (\ref{tauc}) is 
\begin{equation}
\label{taucG}
\tau_c = \sqrt{\frac{\pi}{2}} \sigma ,
\end{equation}
which is independent of the degree of polarization, as far as it involves just polarization fluctuations. We must emphasize that, in contrast to other approaches such as the polarization time in Ref. \cite{SSKF08}, this $\tau_c$ refers just to polarization fluctuations obtained after removing the deterministic part of the polarization state. 

\section*{Example 2}

As a further example already examined in Ref. \cite{SSKF08} we consider the blackbody unpolarized radiation for which 
\begin{equation}
    \delta \Gamma_{\delta S} (\tau ) \propto \left | \zeta \left ( 4, 1+i \tau/\tau_T \right ) \right |^2 , \quad \tau_T = \frac{\hbar}{K_B T},
\end{equation}
where $\zeta (s,a)$ is the generalized Riemann-Hurwitz zeta function, $K_B$ is the Boltzmann constant, $T$ the absolute temperature and $\hbar$ the reduced Planck's constant. By a simple numerical evaluation we get the polarization coherence time $\tau_c = 0.64 \tau_T$ in good agreement with the polarization time in Ref. \cite{SSKF08} since in this case the field is unpolarized. We have plotted  $\Gamma_{\delta S} (\tau)$ and $\left \langle  | \delta \bm{S} (\Omega) |^2 \right \rangle$ in Figs. 1 and 2 respectively,  as functions of $\tau /\tau_T$ and $\Omega \tau_T$.

\begin{figure}[h]
    \includegraphics[width=8cm]{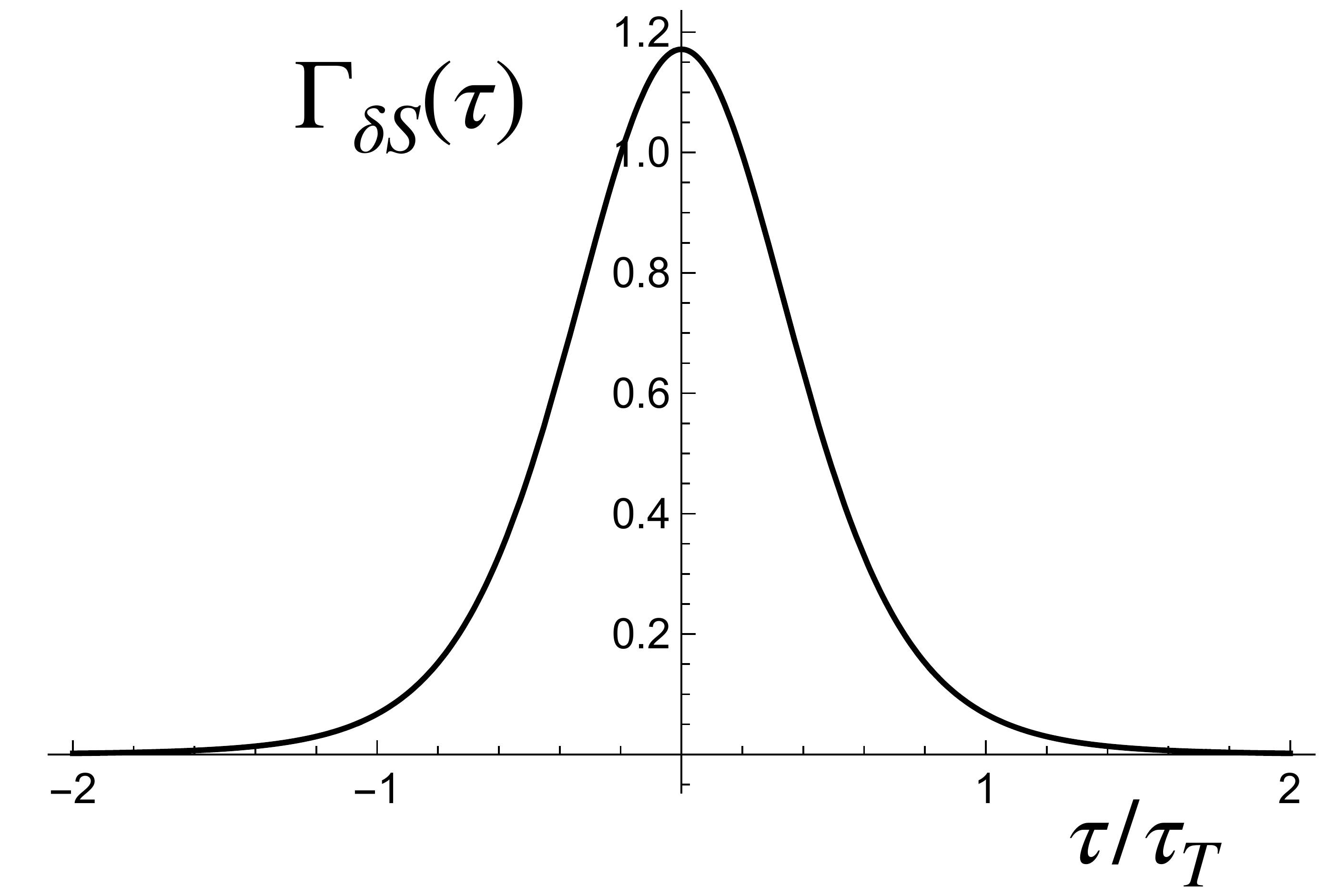}
    \caption{Plot of $\Gamma_{\delta S} (\tau)$ for unpolarized blackbody radiation, as a function of $\tau /\tau_T$.}
\end{figure}{}

\begin{figure}[h]
    \includegraphics[width=8cm]{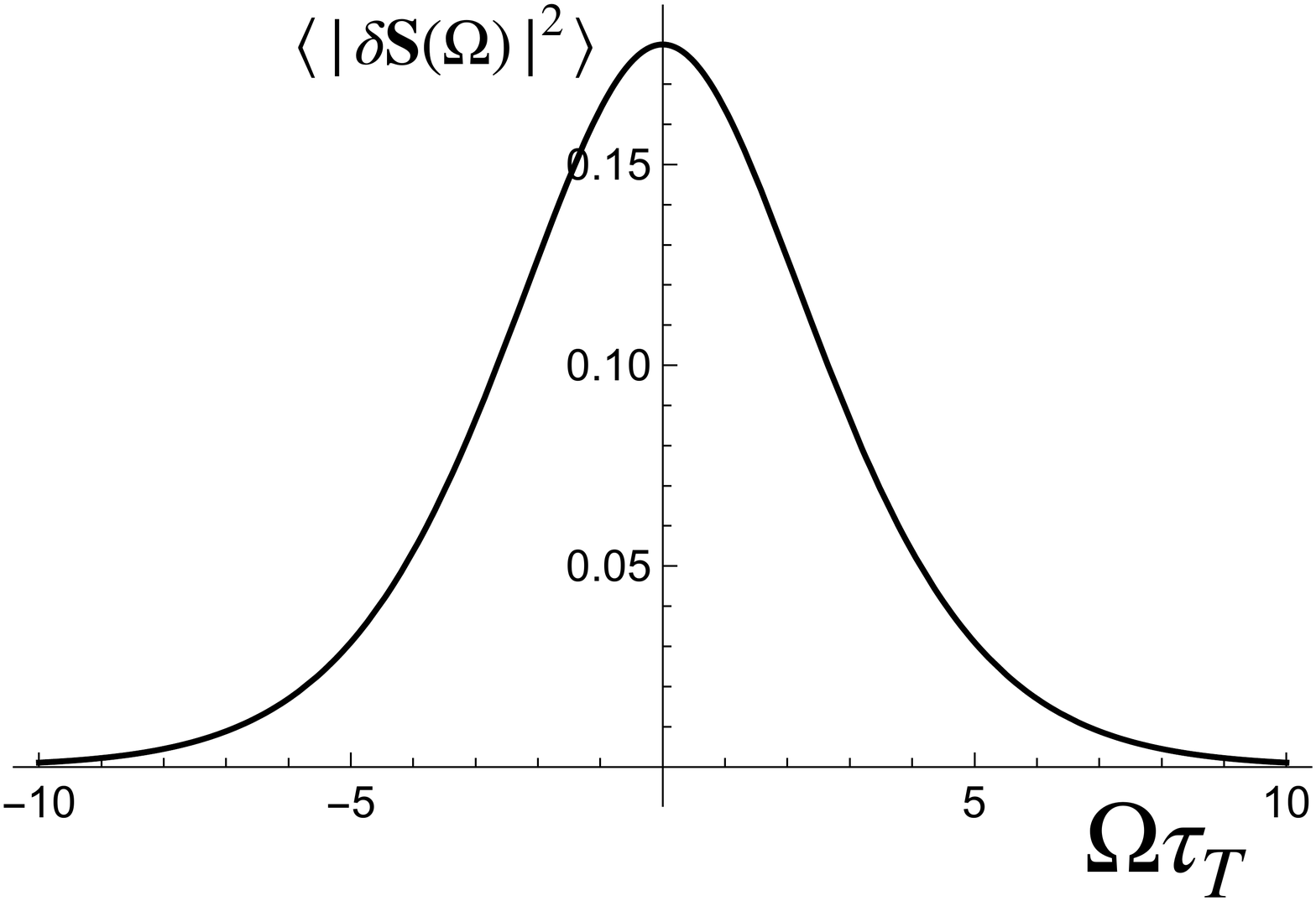}
    \caption{Plot of $\left \langle  | \delta \bm{S} (\Omega) |^2 \right \rangle$ for unpolarized blackbody radiation, as a function of $\Omega \tau_T$.}
\end{figure}{}

\bigskip

\section{Polarization van Cittert-Zernike theorem}

We complete the polarization-coherence analysis addressing the polarization counterpart of another classic coherence theorem. This is the van Cittert-Zernike theorem focusing on the propagation of coherence after an spatially incoherent partially polarized source. Several works have included polarization in the classic van Cittert-Zernike theorem \cite{AJ67,GSBP00,AL07,OMMO09,RT12,TSTF13}. However, these approaches focus on second-order field coherence instead of the four-field polarization correlations addressed here. 

\bigskip

We will work in the space-frequency domain and the frequency value will be thorough removed for the sake of simplicity. The objective is to compute the polarization mutual coherence function (\ref{pcf}) at two spatial points that lie on a plane at a distance $z$ of the source plane, when the source is spatially incoherent, this is
\begin{equation}
    \langle E^\ast_j (\bm{r}_1 ,z=0 ) E_k (\bm{r}_2, z=0 ) \rangle \propto \delta (\bm{r}_1 - \bm{r}_2 ) .
\end{equation}
Moreover we will also assume  Gaussian statistics. For now on $\bm{r}$ denotes the two-dimensional real vectors parametrizing the points on the corresponding plane $z$. We express the field propagation from the source plane to the observation plane in terms of a propagation kernel $K_z(\bm{r},\bm{r}^\prime )$
\begin{equation}
\label{prop}
     E_j (\bm{r} ,z ) = \int d^2 \bm{r}^\prime K_z(\bm{r},\bm{r}^\prime ) E_j (\bm{r}^\prime ,z=0 ) .
\end{equation}

\bigskip

\subsection{Gaussian statistics}

The analysis follows closely the temporal case, so that after recalling that 
\begin{equation}
    \Gamma_S (\bm{r}_1, \bm{r}_2 ,z) = \left \langle  \bm{S} (\bm{r}_1,z) \cdot  \bm{S} (\bm{r}_2 ,z) \right \rangle ,
\end{equation}
we begin with 
\begin{eqnarray}
\label{sums}
  &  \Gamma_S (\bm{r}_1, \bm{r}_2 ,z)  = \sum_{j} \sum_{k, \ell,m,n} \sigma_{k,\ell}^{(j)} \sigma_{m,n}^{(j)} & \nonumber \\ & & \nonumber \\
 & \times \left \langle 
  E^\ast_k (\bm{r}_1,z) E_\ell (\bm{r}_1,z) E^\ast_m (\bm{r}_2,z) E_n (\bm{r}_2,z) \right \rangle & ,
\end{eqnarray}
where $j=1,2,3$ and $k, \ell,m,n=x,y$. Gaussian statistics implies 
\begin{eqnarray}
  &  \Gamma_S (\bm{r}_1, \bm{r}_2 ,z) = \sum_{j} \sum_{k, \ell,m,n} \sigma_{k,\ell}^{(j)} \sigma_{m,n}^{(j)} & \nonumber \\ & & \nonumber \\
 & \times \left [ \left \langle 
  E^\ast_k (\bm{r}_1,z) E_\ell (\bm{r}_1,z) \right \rangle \left \langle E^\ast_m (\bm{r}_2,z) E_n (\bm{r}_2,z) \right \rangle  \right . & \nonumber \\ & & \nonumber \\
  &+ \left . \left \langle 
  E^\ast_k (\bm{r}_1,z) E_n (\bm{r}_2,z)  \right \rangle \left \langle E^\ast_m (\bm{r}_2,z)  E_\ell (\bm{r}_1,z)\right \rangle \right ] . & 
\end{eqnarray}
In the first contribution we can recognize the product Stokes-parameters vectors
\begin{equation}
     \langle \bm{S} (\bm{r}_1,z) \rangle \cdot  \langle \bm{S}   (\bm{r}_2 ,z ) \rangle ,
\end{equation}
that we will embody within the idea of polarization fluctuations in passing from $\Gamma_S$ to $\Gamma_{\delta S}$. Regarding the second term we use again the the Pauli-matrices property (\ref{Pmp}) to get 
\begin{eqnarray}
  &   \Gamma_{\delta S} (\bm{r}_1, \bm{r}_2 ,z) =  & \nonumber \\ & & \nonumber \\
 &  2 \sum_{k, \ell} \left \langle 
  E^\ast_k (\bm{r}_1,z) E_k (\bm{r}_2,z) \right \rangle \left \langle E^\ast_\ell (\bm{r}_2,z) E_\ell (\bm{r}_1,z) \right \rangle  & \nonumber \\ & & \nonumber \\
  &- \sum_{k, m}  \left \langle 
  E^\ast_k (\bm{r}_1,z) E_m (\bm{r}_2,z)  \right \rangle \left \langle E^\ast_m (\bm{r}_2,z)  E_k (\bm{r}_1,z)\right \rangle  , &  \nonumber \\ & & 
\end{eqnarray}
this is 
\begin{eqnarray}
\label{GdS}
& \Gamma_{\delta S} (\bm{r}_1, \bm{r}_2, z) = 2 \left |\mathrm{tr} \bm{W}_z (\bm{r}_1, \bm{r}_2 ) \right |^2 & \nonumber \\
& & \nonumber \\
& - \mathrm{tr} \left [ \bm{W}_z (\bm{r}_1, \bm{r}_2 ) \bm{W}^\dagger_z (\bm{r}_1, \bm{r}_2 )\right ] ,
\end{eqnarray}
where $\bm{W}_z (\bm{r}, \bm{r}^\prime)$ is the cross-spectral density matrix \cite{TSF04}
\begin{equation}
\label{csdm}
\bm{W}_z (\bm{r}_1, \bm{r}_2) =  \left \langle \bm{E}^\ast (\bm{r}_1,z) \bm{E}^T (\bm{r}_2 ,z) \right \rangle ,
\end{equation}
with essentially the same meaning for the two factors in Eq. (\ref{GdS}) than in Eq. (\ref{GdSt}).

\bigskip

\subsection{Gaussian spatially incoherent source}

Finally, under the conditions of the above theorem and assuming the usual Fresnel paraxial approximation, we have the following relations between coherence matrices in the observation and source planes 
\begin{equation}
\bm{W}_z (\bm{r}_1, \bm{r}_2) =  \int d^2 \bm{r}^\prime  K_z (\Delta \bm{r}, \bm{r}^\prime ) \bm{W}_0 (\bm{r}^\prime, \bm{r}^\prime) ,
\end{equation}
where the propagation kernel takes the form
\begin{equation}
K_z (\Delta \bm{r}, \bm{r}^\prime ) = K_z^\ast (\bm{r}_1,\bm{r}^\prime ) K_z (\bm{r}_2,\bm{r}^\prime ) ,
\end{equation}
where $\Delta \bm{r} = \bm{r}_1 - \bm{r}_2$, and
\begin{equation}
\label{KK}
K_z (\Delta \bm{r}, \bm{r}^\prime )\simeq \left (\frac{k}{2 \pi z} \right )^2 e^{-i k \bm{r}^\prime \cdot \Delta \bm{r}/z} ,
\end{equation}
where $k$ is the wave number and we are omitting some phase factors that do not modify the conclusions. With this we have for example 
\begin{equation}
    \mathrm{tr} \left [ \bm{W}_z (\bm{r}_1, \bm{r}_2 ) \right ] = \left (\frac{k}{2 \pi z} \right )^2 \int d^2 \bm{r}^\prime e^{-i \frac{k}{z} \bm{r}^\prime \cdot \Delta \bm{r}} I_0 (\bm{r}^\prime) ,
\end{equation}
where $I_0 (\bm{r}^\prime) = \mathrm{tr} \bm{W}_0 (\bm{r}^\prime, \bm{r}^\prime)$ is the total field intensity at the source point $\bm{r}^\prime$. 

\bigskip

This allows us to express this polarization version of the van Cittert--Zernike theorem in a form similar to the scalar case, this is as a Fourier transform
\begin{equation}
    \Gamma_{\delta S} (\Delta \bm{r}, z) = \left (\frac{k}{2 \pi z} \right )^2 \int d^2 \bm{r}^\prime 
    e^{-i k \bm{r}^\prime \cdot \Delta \bm{r}/z} \mathcal{I} (\bm{r}^\prime) ,
\end{equation}
where 
\begin{eqnarray}
   & \mathcal{I} (\bm{r}) =  \int d^2 \bm{r}^\prime \left \{ 2 I_0 ( \bm{r}^\prime) I_0 ( \bm{r}^\prime-  \bm{r} )  \right .& \nonumber \\ & & \nonumber \\
   & \left . -  \mathrm{tr} \left [ \bm{W}_0 ( \bm{r}^\prime,\bm{r}^\prime) \bm{W}_0 ( \bm{r}^\prime- \bm{r}, \bm{r}^\prime- \bm{r} ) \right ] \right \} .&
   \end{eqnarray}

\bigskip

The essence of the polarization coherence function has been already addressed in Refs. \cite{SPBS21,WHT23}, although their results are not aimed to get to the final form pursued here. In particular, Ref. \cite{SPBS21}  goes beyond scalar coherence functions to consider a $4 \times 4$ coherency matrix, involving all pairs of products of Stokes-parameters fluctuations, and is Fresnel propagation within the Gaussian regime, that certainly offers a perspective worth to be followed. The diagonal elements of such $4 \times 4$ coherency matrix are considered as well in Ref. \cite{WHT23}, always within the Gaussian regime, introducing their spatial power spectral densities.

\subsection{Two-point polarization}

As in the time domain, we may express the result in an slightly different form, 
\begin{equation}
    \Gamma_{\delta S}  (\bm{r}_1, \bm{r}_2 ,z) = \frac{1}{2} \left [ 3- \mathcal{P}^2 (\bm{r}_1, \bm{r}_2,z )\right ] \left |\mathcal{S}_0 (\bm{r}_1, \bm{r}_2 ,z) \right |^2 ,
\end{equation}
where $\mathcal{\bm{S}}$ are the two-point Stokes parameters  \cite{ED04,KW05}
\begin{equation}
\label{tpSp}
    \mathcal{S}_j (\bm{r}_1, \bm{r}_2 ,z ) = \mathrm{tr} \left [ \bm{W}_z (\bm{r}_1, \bm{r}_2 ) \boldsymbol{\sigma}^{(j)} \right ] ,
\end{equation}
and
\begin{equation}
    \mathcal{P} (\bm{r}_1, \bm{r}_2 ,z  )= \frac{|\bm{\mathcal{S}}(\bm{r}_1, \bm{r}_2,z ) |}{|\mathcal{S}_0 (\bm{r}_1, \bm{r}_2 ,z) |} ,
\end{equation}
is a real version of the degree of cross polarization,  where 
\begin{equation}
    \bm{\mathcal{S}}(\bm{r}_1, \bm{r}_2 ,z  ) = \left ( \mathcal{S}_1 (\bm{r}_1, \bm{r}_2 ,z  ), \mathcal{S}_2 (\bm{r}_1, \bm{r}_2 ,z  ), \mathcal{S}_3 (\bm{r}_1, \bm{r}_2 ,z  ) \right )^T .
\end{equation}

\bigskip

\section{Four-dimensional Stokes variables}

It might be interesting to take a quick look at the case where instead of the three-dimensional Stokes vector in Eq. (\ref{3D}) we consider as polarization vector the four-dimensional one
\begin{equation}
\label{4D}
    \tilde{\bm{S}} (\bm{r},t) = \left ( S_0 (\bm{r},t), S_1 (\bm{r},t), S_2 (\bm{r},t), S_3 (\bm{r},t) \right )^T.
\end{equation}
defining a  polarization-fluctuations mutual coherence function
\begin{equation}
    \Gamma_{\delta \tilde{S}} (\bm{r}_1, \bm{r}_2, t_1, t_2 ) = \left \langle \delta  \tilde{\bm{S}} (\bm{r}_1,t_1) \cdot  \delta \tilde{\bm{S}} (\bm{r}_2,t_2 ) \right \rangle .
\end{equation}

\bigskip

Regarding Gaussian statistics, the main difference with the analysis carried out throughout the paper is that now the relevant summation relation of Pauli matrices, that can be easily demonstrated by direct computation, is
\begin{equation}
\label{}
    \sum_{j=0,1,2,3} \sigma_{k,\ell}^{(j)} \sigma_{m,n}^{(j)} =  2 \delta_{k,n}\delta_{\ell,m} ,
\end{equation}
instead of Eq. (\ref{Pmp}).

\bigskip

In the case of temporal coherence this leads to a nice result  for the equivalent of Eq. (\ref{GdStW}) in the form, omitting for simplicity the spatial dependence, 
\begin{equation}
    \Gamma_{\delta \tilde{S}} (t,t+\tau ) 
     = 2 \left | \mathrm{tr} \bm{W} (t,t+\tau ) \right |^2 = 2 |\mathcal{S}_0 (t, t+\tau)|^2,
\end{equation}
where $\mathcal{S}_0  (t, t+\tau)$ is defined in Eq. (\ref{ttSp}). Moreover, in the stationary case the Wiener--Khintchine theorem for polarization fluctuations holds 
\begin{equation}
    \Gamma_{\delta \tilde{S}} (\tau ) = \int d \Omega
\left \langle  | \delta \tilde{\bm{S}} (\Omega) |^2 \right \rangle e^{- i \Omega \tau} ,
\end{equation}
with 
\begin{equation}
\langle  |  \delta \tilde{\bm{S}} ( \Omega )  |^2 \rangle =2 \int d\omega  I (\omega) I (\omega- \Omega) .
\end{equation}

\bigskip

From a more physical perspective we may say that this establishes a deeper connection of polarization and intensity processes, as already pointed out after Eq. (\ref{pf}) in the sense of recalling the practical definition of Stokes parameters in terms of field intensities. 

\bigskip

Regarding spatial coherence the result in Eq. (\ref{GdS}) becomes just,  omitting for simplicity the time dependence, 
\begin{equation}
\Gamma_{\delta \tilde{S}} (\bm{r}_1, \bm{r}_2,z ) = 2 \left | \mathrm{tr} \bm{W}_z (\bm{r}_1, \bm{r}_2 ) \right |^2  = 2 | \mathcal{S}_0 (\bm{r}_1, \bm{r}_2 ,z) |^2,
\end{equation}
where $\mathcal{S}_0 (\bm{r}_1, \bm{r}_2,z )$ is defined in Eq. (\ref{tpSp}). This is a result already found in Ref. \cite{KV19}. In the case of an incoherent source and Fresnel paraxial propagation
\begin{equation}
\label{Gc}
   \Gamma_{\delta \tilde{S}} (\Delta \bm{r},z) = \left (\frac{k}{2 \pi z} \right )^2 \int d^2 \bm{r}^\prime 
    e^{-i k \bm{r}^\prime \cdot \Delta \bm{r}/z} \tilde{\mathcal{I}}(\bm{r}^\prime) ,
\end{equation}
where 
\begin{equation}
 \tilde{\mathcal{I}} (\bm{r}) =2 \int d^2 \bm{r}^\prime I_0 ( \bm{r}^\prime) I_0 ( \bm{r}^\prime-  \bm{r} ) .
\end{equation}

\bigskip

\subsection{Beyond Gaussian}

We may take advantage of the simplicity of the four-dimensional case to address an example beyond the Gaussian realm. Starting in Eq. (\ref{sums}) we will just assume spatial incoherence at the source so that, at $z=0$
\begin{equation}
 \left \langle 
  E^\ast_k (\bm{\rho}_1) E_\ell (\bm{\rho}_1^\prime,z) E^\ast_m (\bm{\rho}_2,z) E_n (\bm{\rho}^\prime_2,z) \right \rangle  
\end{equation}
is different from zero only when 
\begin{equation}
    \bm{\rho}_1=\bm{\rho}_1^\prime, \quad \bm{\rho}_2=\bm{\rho}_2^\prime ,
\end{equation}
as well as when 
\begin{equation}
    \bm{\rho}_1=\bm{\rho}_2^\prime, \quad \bm{\rho}_2=\bm{\rho}_1^\prime .
\end{equation}
Starting with  
\begin{equation}
    \Gamma_{\tilde{S}} (\bm{r}_1, \bm{r}_2) =  \left \langle  \tilde{\bm{S}} (\bm{r}_1)  \cdot  \tilde{\bm{S}} (\bm{r}_2 ) \right \rangle ,
\end{equation}
and  following the same procedure above we get to 
\begin{eqnarray}
\label{nGc}
   & \Gamma_{\tilde{S}} (\Delta \bm{r},z) = 2 \left (\frac{k}{2 \pi z} \right )^2 \int d^2 \bm{\rho}_1 d^2 \bm{\rho}_2 \left \{ \right. & \nonumber \\
    & & \nonumber \\
&    \left \langle  \mathrm{tr} \left [\bm{\mathcal{W}}_0 ( \bm{\rho}_1, \bm{\rho}_1)\bm{\mathcal{W}}_0 ( \bm{\rho}_2, \bm{\rho}_2) \right ] \right \rangle 
    &  \\
      & & \nonumber \\
& + \left . e^{-i k \ \Delta \bm{\rho} \cdot \Delta \bm{r}/z} \left \langle  
\mathrm{tr} \left [ \bm{\mathcal{W}}_0 ( \bm{\rho}_1, \bm{\rho}_2)\bm{\mathcal{W}}^\dagger_0 ( \bm{\rho}_1, \bm{\rho}_2) \right ] \right \rangle 
   \right \} , & \nonumber
\end{eqnarray}
where $\Delta \bm{r} = \bm{r}_1 - \bm{r}_2$, and  $\Delta \bm{\rho} = \bm{\rho}_1 - \bm{\rho}_2$, and $\bm{\mathcal{W}}_z (\bm{\rho}, \bm{\rho}^\prime)$ is the cross-spectral density matrix considered as a field variable, this is before statistical evaluation 
\begin{equation}
\bm{\mathcal{W}}_z (\bm{\rho}, \bm{\rho}^\prime) =   \bm{E}^\ast (\bm{\rho},z) \bm{E}^T (\bm{\rho}^\prime ,z)  ,
\end{equation}
this is that in Eq. (\ref{csdm})
\begin{equation}
\bm{W}_z (\bm{\rho}, \bm{\rho}^\prime) =  \left \langle \bm{\mathcal{W}}_z (\bm{\rho},\bm{\rho}^\prime)\right \rangle .
\end{equation}
To address coherence for polarization fluctuations we must as well consider the term
\begin{equation}
    \left \langle   \tilde{\bm{S}} (\bm{r}_1,t_1) \right \rangle \cdot  \left \langle  \tilde{\bm{S}} (\bm{r}_2,t_2 ) \right \rangle  ,
    \end{equation}
which leads to 
\begin{eqnarray}
\label{ptS}
   &\left \langle   \tilde{\bm{S}} (\bm{r}_1,t_1) \right \rangle \cdot  \left \langle  \tilde{\bm{S}} (\bm{r}_2,t_2 ) \right \rangle  = 2 \left (\frac{k}{2 \pi z} \right )^2 \int d^2 \bm{\rho}_1 d^2 \bm{\rho}_2 \left \{ \right. & \nonumber \\
    & & \nonumber \\
&   \mathrm{tr} \left [\bm{W}_0 ( \bm{\rho}_1, \bm{\rho}_1)\bm{W}_0 ( \bm{\rho}_2, \bm{\rho}_2) \right ]
    &  \\
      & & \nonumber \\
& + \left . e^{-i k \ \Delta \bm{\rho} \cdot \Delta \bm{r}/z} 
\mathrm{tr} \left [ \bm{W}_0 ( \bm{\rho}_1, \bm{\rho}_2)\bm{W}^\dagger_0 ( \bm{\rho}_1, \bm{\rho}_2) \right ] 
   \right \} . & \nonumber
\end{eqnarray}
In comparison with the Gaussian case in Eq (\ref{Gc}) we may notice that the Fourier-transform relation is partially lost in general because of the firsts factors in Eqs. (\ref{nGc}) and (\ref{ptS}). Furthermore the factor in the Fourier transform is very close to the Gaussian case in the sense that we have the equality 
\begin{eqnarray}
   & \mathrm{tr} \left [ \bm{\mathcal{W}}_0 ( \bm{\rho}_1, \bm{\rho}_2)\bm{\mathcal{W}}^\dagger_0 ( \bm{\rho}_1, \bm{\rho}_2) \right ]= & \nonumber \\ &  & \nonumber \\  &  \mathrm{tr} \left [ \bm{\mathcal{W}}_0 ( \bm{\rho}_1, \bm{\rho}_1) \right ]   \mathrm{tr} \left [ \bm{\mathcal{W}}_0 ( \bm{\rho}_2, \bm{\rho}_2) \right ] &,
\end{eqnarray}
so that the two Fourier transforms in Eqs. (\ref{Gc}) and (\ref{nGc}) only differ in the  replacement of 
\begin{equation}
   \left \langle  \mathrm{tr} \left [ \bm{\mathcal{W}}_0 ( \bm{\rho}_1, \bm{\rho}_1)\bm{\mathcal{W}}_0 ( \bm{\rho}_2, \bm{\rho}_2) \right ] \right \rangle ,
\end{equation}
by 
\begin{equation}
\left \langle  \mathrm{tr} \left [ \bm{\mathcal{W}}_0 ( \bm{\rho}_1, \bm{\rho}_1) \right ] \right \rangle  \left \langle  \mathrm{tr} \left [ \bm{\mathcal{W}}_0 ( \bm{\rho}_2, \bm{\rho}_2) \right ] \right \rangle .
\end{equation}

\section{Conclusions}

We have studied suitable polarization-coherence versions of two celebrated theorems in classical-optics coherence. These are the Wiener-Khintchine and van Cittert-Zernike theorems dealing with the time-frequency and spatial manifestations of coherence. This can be properly done via a suitable polarization-fluctuation coherence function. 

The coherence theorems we have developed are fully expressed in terms of four-field correlations as it corresponds to correlations of polarization variables properly defined in terms of Stokes variables. Given the power and usefulness of the classic coherence theorems we think this polarization version is quite worth investigating. This has allowed us to introduce a suitable definition of polarization time entering in an exact duality relation with polarization spectral width. We have also derived suitable extensions of the theorem to four-dimensional Stokes parameters with extremely simple results that may shed light to the problem of coherence for polarized light.

\bigskip

\section*{ACKNOWLEDGMENTS} 
A. L. acknowledges financial support from project PR44/21--29926 from Santander Bank and Universidad Complutense of Madrid.

\end{document}